\documentclass[preprint,proceedings]{rmaa}

\SetYear{2003}
\SetConfTitle{Compact Binaries in the Galaxy and
Beyond}

\title{Thermal Radiation from an Accretion Disk}

\author{F.V.Prigara \altaffilmark{1}}

\altaffiltext{1}{Institute of Microelectronics and Informatics,
Russian Academy of Sciences, 21 Universitetskaya, Yaroslavl
150007, Russia(\protect\email{fprigara@imras.yar.ru}).}

\suppressfulladdresses

\listofauthors{F.V.Prigara}

\indexauthor{Prigara, F.V.}

\addkeyword{accretion, accretion disks}
\addkeyword{radiation
mechanisms: general}

\begin{document}

\maketitle

\boldabstract{An effect of stimulated radiation processes on
thermal radiation from an accretion disk is considered. The radial
density waves triggering flare emission and producing
quasi-periodic oscillations in radiation from an accretion disk
are discussed. It is argued that the observational data suggest
the existence of the weak laser sources in a two-temperature
plasma of an accretion disk.}

It was shown earlier that thermal radio emission has a stimulated
character and that this statement is probably valid for thermal
blackbody radiation in others spectral ranges (Prigara 2003).
According to this conception thermal radiation from non-uniform
gas is produced by an ensemble of individual emitters. Each of
these emitters is an elementary resonator the size of which has an
order of magnitude of mean free path of photons, $l = 1/\left(
{n\sigma} \right)$, where \textit{n} is the number density of
particles, and $\sigma $ is the absorption cross-section.

The emission of each elementary resonator is coherent, with the
wavelength $\lambda = l$, and thermal radiation from a gaseous
layer is incoherent sum of radiation produced by individual
emitters. This dependence of the wavelength on the density is
consistent with the experimental results by Looney and Brown on
the excitation of plasma oscillations (Alexeev 2003).

In the gaseous disk model (Prigara 2003), the density depends upon
the distance \textit{r} from the energy center as follows

\begin{equation}
\label{eq1}
n \propto \left( {v_{0}^{2} r^{2} + c^{2}r_{s} r} \right)^{ - 1/2},
\end{equation}

\noindent where $r_{s} $ is the Schwarzschild radius, \textit{c}
is the speed of light, and $v_{0} $ is a constant. This density
profile is consistent with the convection dominated accretion flow
(CDAF) models or the advection dominated accretion flow (ADAF)
models with an outflow (Nagar et al. 2001).

The last equation gives the wavelength dependence of the emitting
region size $r_{\lambda}  $ in the form $r_{\lambda}  \propto
\lambda ^{2}$ at small values of the radius, and $r_{\lambda}
\propto \lambda $ at the large radii. These relations are indeed
observed in radio band for compact and extended radio sources
respectively. Compare these results with the Lyndell-Bell's
relation $r_{\lambda} \propto \lambda ^{4/3}$ .

Some X-ray binaries, e.g. Sco X-1, show weak emission lines with
the variable intensities and radial velocities. These features are
characteristic for non-saturated lasers. The wavelength of
generated mode is determined by the size of an elementary
resonator which is depending on the density. The radial density
wave travelling along the radius changes the density which results
in the variations of the wavelength. Convection or advection in
the gaseous disk produces a two-temperature plasma in which it is
possible to create the inversion of the energy level population.
This gives rise to the weak lasers similar to the weak molecular
masers.

The radial density wave produces also the delay of flares at low
frequencies and high frequency quasi-periodic oscillations
observed in X-ray binaries. Density waves can presumably propagate
in a sufficiently hot plasma of an accretion disk. They differ
from the shocks, since the density waves have the smooth density
and pressure profiles.

The density waves in a hot plasma seem to have been observed in a
solar microwave burst, causing the strong correlation in the
modulation of radio and X-ray emission over a large distance
($10^{10}cm$) in the solar corona (Grechnev et al. 2003).

\end{document}